\newcounter{myctr}
\def\myitem{\refstepcounter{myctr}\bibfont\noindent\ifnum\themyctr>9\else\phantom{0}\fi\hangindent17pt\themyctr.\enskip}

\documentclass{ws-ijqi}
\usepackage{hyperref}
\usepackage[super,sort,compress]{cite}
\usepackage{url}
\usepackage{color}
\usepackage{comment} 
\usepackage{bbm}
\usepackage{xparse}
\NewDocumentCommand{\ceil}{s O{} m}{%
  \IfBooleanTF{#1} 
    {\left\lceil#3\right\rceil} 
    {#2\lceil#3#2\rceil} 
}
\def\Id{{\mathbbm 1}}
\DeclareMathOperator{\Tr}{Tr}
\def\nl{{\rm (nl)}}
\def\erf{{\rm erf}}
\def\dag{{\dagger}}
\def\inf{{\infty}}
\def\HEL{\mathrm{Hel}}
\def\SQL{\mathrm{SQL}}
\def\HY{\mathrm{HY}}
\def\HD{{\rm HD}}
\def\OUTPUT{\xi}
\def\nth{n_{\rm th}}
\def\disp{{\rm D}}
\def\sigmamax{\sigma_{\rm max}}
\def\opt{{\rm opt}}

\begin{document}

\catchline{}{}{}{}{}

\title{A robust hybrid receiver for binary phase-shift keying discrimination in the presence of phase noise}

\author{Michele N. Notarnicola and Stefano Olivares\footnote{Corresponding author.}}
\address{Dipartimento di Fisica ``Aldo Pontremoli'',\\
Universit\`a degli Studi di Milano, I-20133 Milano, Italy\\[2ex]
INFN, Sezione di Milano, I-20133 Milano, Italy\\
stefano.olivares@fisica.unimi.it}

\maketitle

\begin{history}
\received{\today}
\end{history}

\begin{abstract}
We address the problem of coherent state discrimination in the presence of phase diffusion. We investigate the role of the hybrid near-optimum receiver (HYNORE) we proposed in [{\em J. Opt. Soc. Am. B\/} {\bf 40}, 705-714 (2023)] in the task of mitigating the noise impact. We prove the HYNORE to be a robust receiver, outperforming the displacement photon-number-resolving (DPNR) receiver and beating the standard quantum limit in particular regimes. We introduce the maximum tolerable phase noise $\sigmamax$ as a figure of merit for the receiver robustness and show that HYNORE increases its value with respect to the DPNR receiver.
\end{abstract}


\markboth{M.~N.~Notarnicola \& S.~Olivares}
{A robust hybrid receiver for binary phase-shift keying discrimination$\ldots$}

\section{Introduction}
Quantum discrimination of coherent states is a central problem for quantum technologies, as they represent the 
typical information carriers in both fiber-optics and deep-space communications \cite{Agrawal2002, Kikuchi2016, Kaushal2017}. Within this field, the goal is to design a quantum receiver minimizing the decision error probability \cite{Helstrom1976, Bergou2010, Cariolaro2015}, being a possible resource to enhance both quantum communications \cite{Giovannetti2004, Arrazola2014} and quantum key distribution \cite{Grosshans2002, Gisin2002, Leverrier2009, Notarnicola2023-Pol}.
These kinds of receivers provide a genuine quantum advantage over the standard receivers, based on either homodyne or double-homodyne detection. Indeed, while conventional receivers are limited by the shot noise limit, or standard quantum limit (SQL), the quantum decision theory developed by Helstrom identifies the optimum receiver achieving the minimum error probability allowed by quantum mechanics, namely the Helstrom bound \cite{Helstrom1970}.

The decision problem has been widely studied for binary phase-shift keying (BPSK) discrimination. The optimum receiver
was designed by Dolinar \cite{Dolinar1973, Lau2006} and several proposals of more feasible schemes have been advanced, employing both single-shot \cite{Kennedy1973, Takeoka2008, DiMario2018, Notarnicola2023-HY} and feed-forward strategies \cite{Assalini2011, Sych2016, Notarnicola2023-HFF}. Among them, strategies employing subitable displacement operations like the Kennedy receiver \cite{Kennedy1973} and the hybrid near-optimum receiver (HYNORE) \cite{Notarnicola2023-HY} proved themselves to be near-optimum, beating the SQL and providing useful detection schemes to enhance information transfer over attenuating (Gaussian) channels \cite{Notarnicola2023-Pol}. 

Beside this, the performance of quantum receivers in the presence of noisy non-Gaussian channels is another fundamental task towards the realistic implementation of quantum optical communications.
A paradigmatic example is provided by phase noise\cite{Banerjee2007, Ishii11,Amiri22,Bertaina23}, which represents the most detrimental source of noise for phase-shift encoding, destroying the coherence and the purity of the employed coherent pulses \cite{Cialdi2020, Genoni2011}.
The impact of phase diffusion has been investigated in quantum metrology \cite{Genoni2011, Altorio2015, Szczykulska2017, Notarnicola2022}, quantum interferometry \cite{Genoni2012, Vidrighin2014, Feng2014}, quantum communications \cite{Jarzyna2014, Trapani2015, Namkung2022} and quantum state discrimination \cite{Olivares2013,Chesi2018,DiMario2019}. Remarkably, the Kennedy receiver is no longer near-optimum for BPSK discrimination over a phase diffusion channel, and homodyne detection approaches the Helstrom bound in the regime of large noise \cite{Olivares2013}. To overcome this fundamental limitation, DiMario and Becerra adopted the displacement photon-number-resolving (DPNR) receiver \cite{DiMario2018}, that is a Kennedy setup employing photon-number-resolving (PNR) detectors with finite resolution $M$ instead of on-off photodetectors, and optimized also the encoding strategy, enhancing robustness to both channel and detection noise \cite{DiMario2019}.

In this paper we investigate the performance of the HYNORE \cite{Notarnicola2023-HY} in the presence of phase diffusion. The HYNORE is a hybrid receiver that combines the setups of weak-field homodyne, or homodyne-like (HL), detection \cite{Donati2014,Thekkadath2020} and the DPNR receiver. Its basic principle is to split the encoded signal at a beam splitter, implement HL detection on the reflected branch and, according to the obtained result, perform a suitable displacement operation on the transmitted side. Thereafter, in ideal conditions it behaves as a near-optimum receiver beating both the DPNR receiver and the SQL, and also shows more robustness to detection noise and visibility reduction \cite{Notarnicola2023-HY, Notarnicola2023-HFF}. Here, we address its role to mitigate the detriments of phase diffusion in BPSK discrimination and show it to outperform the DPNR receiver in different regimes, in particular in the high-energy and large-noise limits.

The structure of the paper is the following. In Sec.~\ref{sec: PhN} we briefly recall the main features of coherent state discrimination, while Sec.~\ref{sec: DPNR} deals with the problem of discrimination in the presence of phase diffusion, presenting in detail the DPNR receiver proposed in Ref.s \citen{DiMario2018, DiMario2019}. Then, Sec.~\ref{sec: HYNORE} addresses the performance of HYNORE for phase diffused coherent states and proves it to outperform the DPNR scheme. Finally, in Sec.~\ref{sec: Concl} we draw some conclusions.

\section{Basics of coherent state discrimination}\label{sec: PhN}

In the conventional BSPK encoding in the absence of phase diffusion, the goal is to perform discrimination between two symbols $k=0,1$, encoded into the two coherent states \cite{Cariolaro2015}:
\begin{align}
|\alpha_k\rangle =|e^{i\pi (k+1)} \, \alpha \rangle \, ,
\end{align}
with $\alpha>0$, $\alpha^2$ being the mean energy, generated by a quantum source with equal a priori probabilities.
To this aim, we shall design a quantum receiver, described by a binary positive-operator valued measurement (POVM) $\{\Pi_0, \Pi_1\}$, associated with the error probability $P_{\rm err}=[P(0|1)+P(1|0)]/2$, where $P(j|k)= \langle\alpha_k|\Pi_j |\alpha_k\rangle$, $j,k=0,1$, is the probability of inferring symbol $j$ if the state $|\alpha_k\rangle$ was measured.

The minimum error probability allowed by quantum mechanics, namely the Helstrom bound \cite{Helstrom1976, Bergou2010, Cariolaro2015}, in the noiseless case is equal to $P_\HEL^\nl=[1-\sqrt{1-\exp(-4\alpha^2)}]/2$, and it is achieved by the Dolinar receiver \cite{Dolinar1973}. Nevertheless, the Dolinar setup requires optical feedback and continuous-time measurements, thus making its practical implementation quite challenging. 
On the other hand, standard receivers reach the SQL, obtained with homodyne detection, with the associated error probability $P_\SQL^\nl=[1-\erf(\sqrt{2}\alpha)]/2$. Given this scenario, the task of quantum state discrimination theory is to design a feasible receiver outperforming the SQL and being as close as possible to the Helstrom bound. 
In the absence of phase diffusion, two paradigmatic examples of feasible receivers are provided by the Kennedy receiver \cite{Kennedy1973, DiMario2018} and the HYNORE \cite{Notarnicola2023-HY}.

In the Kennedy receiver, or displacement receiver, the incoming signal $|\alpha_k\rangle$ undergoes the displacement operation \cite{Olivares2021} $D(\alpha)$, followed by on-off detection.
The displacement may be implemented practically by letting the signals interfere with a suitable intense local oscillator at a beam splitter with large transmissivity \cite{Paris1996}.
We note that $D(\alpha)$ performs a nulling operation, leading to the mapping:
\begin{align}
|-\alpha\rangle \rightarrow |0\rangle \quad \mbox{and} \quad |\alpha\rangle \rightarrow |2\alpha\rangle \, ,
\end{align}
and, in particular, sending $|\alpha_0\rangle$ into the vacuum. Therefore, BPSK is turned into on-off keying (OOK) and on-off detection provides the optimal measurement choice. The resulting error probability is equal to $P_{\disp}^\nl=\exp(-4\alpha^2)/2$, proving the receiver to be near-optimum, namely proportional to the Helstrom bound, in the high-energy limit $\alpha^2\gg 1$. Moreover, it also beats the SQL for $\alpha^2 > \alpha^2_{\disp}$, with $\alpha^2_\disp\approx 0.38$ \cite{Kennedy1973, Notarnicola2023-HY}.

On the other hand, the HYNORE is a hybrid near-optimum receiver based on the combination of the Kennedy setup and the weak-field homodyne, or HL, detection. Its functioning will be briefly presented in the next subsection.

\subsection{The hybrid near-optimum receiver}
\begin{figure}[t]
\centerline{\includegraphics[width=0.8\textwidth]{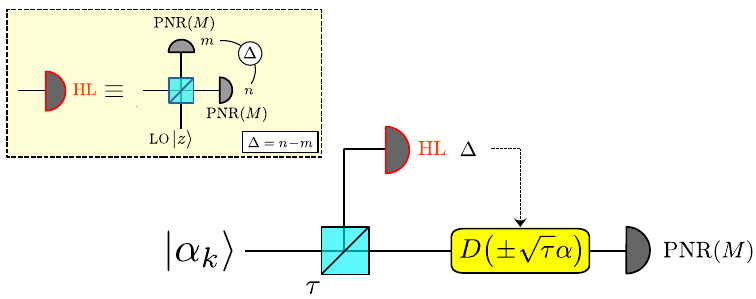}}
\caption{Scheme of the HYNORE. The incoming signal is split at a beam splitter with transmissivity $\tau$, thereafter HL detection is implemented on the reflected branch. The outcome $\Delta = n - m$ is exploited to decide the displacement operation implemented on the transmitted signal.
(Inset) Setup of the weak-field homodyne, or homodyne-like, detection. The signal is mixed at a balanced beam splitter with a low-intensity local oscillator (LO), and PNR$(M)$ detection is performed on the output modes.}\label{fig:01-HY-scheme}
\end{figure}
The HYNORE employs HL detection, that is a homodyne setup where the conventional p-i-n photodiodes and high-intensity local oscillator (LO) are replaced with PNR detectors and low-intensity LO, respectively \cite{Donati2014, Thekkadath2020}. Its implementation is sketched in the inset of Fig.~\ref{fig:01-HY-scheme}. First, the incoming signal impinges at a balanced beam splitter with a low-intensity LO $|z\rangle$, $z>0$; thereafter we implement PNR detection on both the output modes, retrieving the outcomes $n$ and $m$, and finally we consider the difference photocurrent $\Delta=n-m$. 

Realistic PNR detectors have a finite resolution $M$, as they can resolve up to $M$ photons. In turn, they are described by the POVM $\{\Pi_0, \Pi_1, \ldots, \Pi_M \}$ with $M+1$ elements, where:
\begin{align}
\Pi_n = 
\left\{\begin{array}{l l} 
 | n \rangle\langle n| &\mbox{if}~ n=0,\ldots, M-1 \, , \\[2ex]
 \displaystyle \Id - \sum_{j=0}^{M-1} | j \rangle\langle j| \quad &\mbox{if}~ n=M .
\end{array}
\right.
\end{align}
For a better clarity, from now on we refer to them as PNR$(M)$ detectors. In particular, PNR$(1)$ corresponds to on-off detection, whereas ideal photodetectors are referred to as PNR$(\inf)$.
Accordingly, we have $n,m=0,\ldots, M$ and the difference photocurrent gets only integer values in the range $-M\le \Delta\le M$.
For an input coherent signal $|\gamma\rangle$, $\gamma\in \mathbb{C}$, the HL probability distribution reads\cite{Notarnicola2023-HY}:
\begin{align}\label{eq:HL}
{\cal S}_\Delta(\gamma) = \sum_{n,m=0}^{M} q_n\left(\mu_{+}(\gamma)\right) \, q_m\left(\mu_{-}(\gamma)\right) \,  \delta_{(n-m),\Delta} \, ,
\end{align}
where
\begin{align}
\mu_{\pm}(\gamma)= \frac{|\gamma\pm z|^2}{2} \, ,
\end{align}
are the mean number of photons, or energies, on the two output branches, $\delta_{jk}$ is the Kronecker Delta, and
\begin{align}\label{eq:qn}
    q_n(\mu) =
    \left\{\begin{array}{l l}
    {\displaystyle e^{-\mu} \ \frac{\mu^{n}}{n!}}  & \mbox{if}~n<M \ , \\[2ex]
    {\displaystyle 1- e^{-\mu} \sum_{j=0}^{M-1} \frac{\mu^{j}}{j!}} \quad & \mbox{if}~n = M \, ,
    \end{array}
    \right.
\end{align}
is the probability of getting outcome $n$ after PNR$(M)$ detection \cite{Notarnicola2023-HY, Notarnicola2023-HFF, DiMario2018}.

The HL setup may be suitably exploited to design a hybrid receiver, the HYNORE, improving the performance of the Kennedy, whose structure is reported in Fig.~\ref{fig:01-HY-scheme}. The underlying principle is to implement HL on a portion of the encoded signal to improve the displacement operation of the Kennedy scheme. Both the fraction of the signal directed to the HL detector and the amplitude of the LO are optimized to minimize the overall error probability. 
In more detail, the state $|\alpha_k\rangle$ is split at a beam splitter of transmissivity $\tau$, such that
\begin{align}\label{eq:alphaBS}
|\alpha_k\rangle \rightarrow |\alpha_k^{(r)}\rangle \otimes |\alpha_k^{(t)}\rangle \equiv |-\sqrt{1-\tau} \alpha_k\rangle \otimes |\sqrt{\tau}\alpha_k\rangle \, .
\end{align}
Firstly, we perform HL detection on the reflected signal $|\alpha_k^{(r)}\rangle$. The value of the obtained outcome $\Delta$ provides us with some information on the field phase, which can be exploited to choose the sign of a displacement operation $D(\pm\sqrt{\tau}\alpha)$ to be implemented on the transmitted pulse $|\alpha_k^{(t)}\rangle$. Indeed, if $\Delta\ge 0$ it is more likely that state $|\alpha_0\rangle$ was sent, therefore we apply $D(+\sqrt{\tau}\alpha)$, while in the opposite case we choose $D(-\sqrt{\tau}\alpha)$. Eventually, we perform on-off detection, which may be still implemented by PNR$(M)$ detectors. The final decision rule is depicted in Table~\ref{Tab:01-DecRule}.

\begin{table}[ph]
\tbl{Decision strategy for the HYNORE in Fig.~\ref{fig:01-HY-scheme}.\label{Tab:01-DecRule}}
{\begin{tabular}{@{}ccc@{}} \toprule
outcomes & & decision \\  \colrule
    $\Delta \ge 0$ $\quad$ off \, & & ``0" \\ 
    $\Delta < 0$ $\quad$ on \, & & ``0" \\ 
    $\Delta < 0$ $\quad$ off \, & & ``1" \\ 
    $\Delta \ge 0$ $\quad$ on \, & & ``1" \\ \botrule
\end{tabular}}
\end{table}

The error probability then reads:
\begin{align}\label{eq:Phynore}
P_\HY^\nl =\min_{\tau, z} \left\{ \frac{e^{-4 \tau \alpha^2}}{2} \,\left[
\sum_{\Delta=-M}^{-1} {\cal S}_\Delta\left(\alpha^{(r)}_0\right)
+ \sum_{\Delta=0}^{M} {\cal S}_\Delta \left(\alpha^{(r)}_1\right) \right] \right\}\, ,
\end{align}
optimized over the transmissivity $\tau$ and the LO amplitude $z$ of the HL scheme.

As demonstrated in Ref.s~\citen{Notarnicola2023-HY, Notarnicola2023-HFF}, the HYNORE is near-optimum, it outperforms the Kennedy receiver for all energies, $P_\HY^\nl \le P_{\rm K}^\nl$, and beats the SQL for $\alpha^2>\alpha^2_\HY(M)$, where $\alpha^2_\HY(M)$ is a decreasing function of the resolution $M$ such that $\alpha^2_\HY(M)<\alpha^2_\disp$.

\section{Quantum discrimination of phase diffused coherent states}\label{sec: DPNR}
\begin{figure}[t]
\centerline{\includegraphics[width=0.9\textwidth]{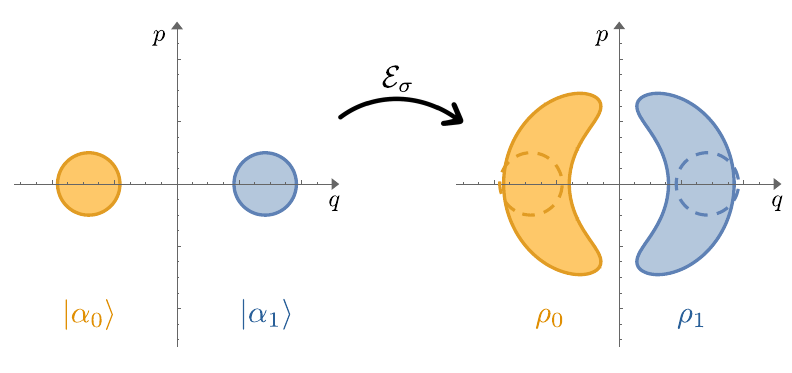}}
\caption{Phase space representation of the BPSK encoding before (left) and after (right) the application of phase diffusion. The phase diffusion CP map $\mathcal{E}_\sigma$ transforms the initial coherent signals into a Gaussian mixture of phase-shifted coherent states, reducing both purity and coherence.}\label{fig:02-PhSp}
\end{figure}

In the presence of a phase diffusion channel, the problem of BPSK discrimination is remarkably different with respect to the scenario discussed in Sec.~\ref{sec: PhN}. In fact, the encoded coherent states evolve according to a suitable master equation \cite{Genoni2011}, equivalent to the completely positive (CP) map $\mathcal{E}_\sigma$, such that:
\begin{align}\label{eq:Map}
|\alpha_k\rangle \xrightarrow[]{\mathcal{E}_\sigma} \rho_k=\int_\mathbb{R}  d\phi \, g_\sigma(\phi) \, | \alpha_k e^{-i\phi} \rangle \langle \alpha_k e^{-i\phi} | \, ,
\end{align}
where $g_\sigma(\phi)= \exp[-\phi^2/(2\sigma^2)]/\sqrt{2\pi \sigma^2}$ is a Gaussian distribution whose standard deviation $\sigma>0$ quantifies the amount of noise.
That is, the overall effect of phase diffusion is the application of a Gaussian-distributed random phase shift to the incoming signal, resulting in a overall non-Gaussian CP map. 

The map (\ref{eq:Map}) not only provides a model for light propagation through phase-fluctuating quantum channels \cite{Ip2008}, but may be also adopted in other contexts. An intriguing case is local-local oscillator continuous-variable quantum key distribution (LLO-CVQKD), where a sender and a receiver share a reference phase via quantum estimation on a coherent probe \cite{Marie2017, Chin2021, Shao2021}. The phase value is estimated after double-homodyne detection, and the obtained result is assumed to fluctuate around the actual value according to a Gaussian distribution. Another example is represented by modeling the phase fluctuations of a realistic laser source \cite{Cialdi2020, Notarnicola2022}. More precisely, phase noise of laser radiation is a diffusive stochastic process characterized by both a phase drift and phase fluctuations. While the drift can be controlled by monitoring part of the radiation emitted by the source \cite{Bina2016}, mitigating phase fluctuations is a more challenging task \cite{Cialdi2020, Notarnicola2022} and, in principle, they shall be taken into account in realistic quantum protocols.

Given the previous considerations, in the presence of BPSK the effect of phase diffusion is detrimental: it reduces both the coherence and the purity of the encoded coherent states as emerges from Fig.~\ref{fig:02-PhSp}, reporting the phase space representation of the quantum states before and after the noisy channel.
In turn, the quantum receiver has to discriminate between the two mixed phase-diffused states $\rho_0$ and $\rho_1$. The Helstrom bound becomes \cite{Helstrom1976, Bergou2010, Cariolaro2015}:
\begin{align}\label{eq:HEL}
P_\HEL= \frac12 \left[1- \frac12\Tr \big( |\Lambda | \big) \right] \, ,
\end{align}
with $\Lambda=  \rho_0 -  \rho_1$.
The corresponding SQL reads \cite{Olivares2013}:
\begin{align}
P_{\SQL} = \frac12 \left[ \int_0^\inf dx \, p_\HD (x|0) + \int_{-\inf}^{0} dx \, p_\HD(x|1)\right] \, ,
\end{align}
where
\begin{align}
p_\HD(x|k) = \int_{\mathbb{R}} d\phi \, g_\sigma(\phi) \, \frac{\exp\left[-(x-2\alpha_k \cos\phi)^2/2\right]}{\sqrt{2\pi}}
\end{align}
is the homodyne probability of obtaining outcome $x$ given the state $| \alpha_k \rangle$, expressed in shot-noise units.

As regards the Kennedy receiver, the presence of phase noise is detrimental and its performance is severely degraded for $\sigma>0$ \cite{Olivares2013}. To mitigate the noise impact, DiMario and Becerra proposed the DPNR receiver \cite{DiMario2018, DiMario2019}, namely a Kennedy setup where PNR($M$) detectors replace the on-off ones. 
In fact, for $\sigma> 0$  the displacement operation $D(\alpha)$ maps states $\rho_k$ into $\OUTPUT_k= D(\alpha) \rho_k D(\alpha)^\dag$, equal to:
\begin{align}\label{eq:OutStates}
\OUTPUT_k&= \int_\mathbb{R} d\phi \, g_\sigma(\phi) \, \left|\sqrt{\mu_k(\alpha^2,\phi)} \, e^{-i\phi/2}\right\rangle \left\langle \sqrt{\mu_k(\alpha^2,\phi)} \, e^{-i\phi/2} \right| \, ,
\end{align}
where
\begin{align}\label{eq:muk}
\mu_0(\alpha^2,\phi) = 4\alpha^2 \sin^2(\phi/2) \quad \mbox{and} \quad \mu_1(\alpha^2,\phi) = 4\alpha^2 \cos^2(\phi/2) \, .
\end{align}
Differently from the noiseless case, the nulling operation implemented by $D(\alpha)$ is not perfect and the output state $\OUTPUT_0$ still contains some photons.
Thereby, on-off detection is not the most appropriate strategy anymore and the DPNR setup is expected to outperform the Kennedy.

The PNR$(M)$ probability distribution of the displaced states $\OUTPUT_k$ reads:
\begin{align}
p_\sigma(n|k) = \int_\mathbb{R} d\phi \, g_\sigma(\phi) \, q_n\left(\mu_k(\alpha^2,\phi) \right) \, , \quad (n=0,\ldots,M) \, ,
\end{align}
with the probability $q_n$ and the count rates  $\mu_k(\alpha^2,\phi)$ defined in Eq.s~(\ref{eq:qn}) and~(\ref{eq:muk}), respectively.
The final decision is performed according to the maximum a posteriori probability (MAP) criterion:  given the outcome $n=0,\ldots,M$, we infer the state “0” or “1” associated with the maximum a posteriori probability \cite{DiMario2018, DiMario2019}. This is equivalent to introducing a threshold $\nth= \nth(\alpha^2,\sigma)\le M$ such that all outcomes $n < \nth$ correspond to decision ``0", while outcomes $n \ge \nth$ infer state ``1" \cite{Notarnicola2023-HY}. The threshold is obtained numerically by equating the photon number distributions of the two displaced phase-diffused states, namely $p_\sigma(\bar{n}|0)=p_\sigma(\bar{n}|1)$, $\bar{n} \in {\mathbbm R}$, and considering the lowest integer greater than the obtained root $\bar{n}$, namely $\nth(\alpha^2,\sigma)=\ceil{\bar{n}}$.
In turn, the error probability of the DPNR receiver reads:
\begin{align}
P^{(M)}_\disp= \frac12 \left[ \sum_{n=0}^{\nth-1} p_\sigma(n|1) + \sum_{n=\nth}^{M} p_\sigma(n|0)\right] \, ,
\end{align}
and with PNR$(1)$ detection we retrieve the Kennedy receiver.

\begin{figure}[tb!]
\centerline{
\includegraphics[width=0.5\textwidth]{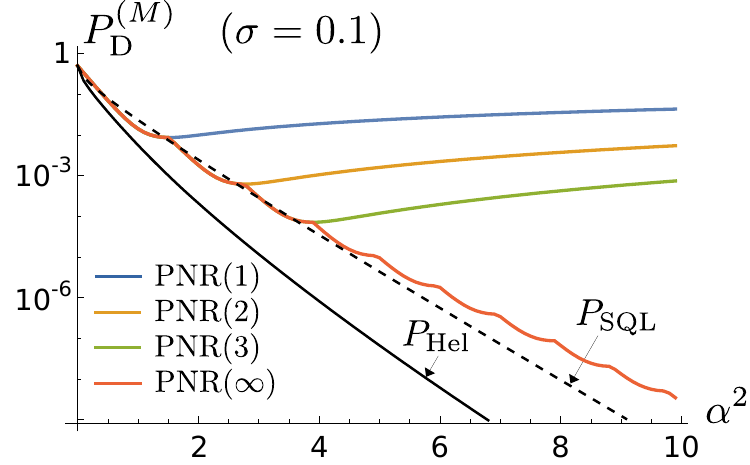}
\includegraphics[width=0.5\textwidth]{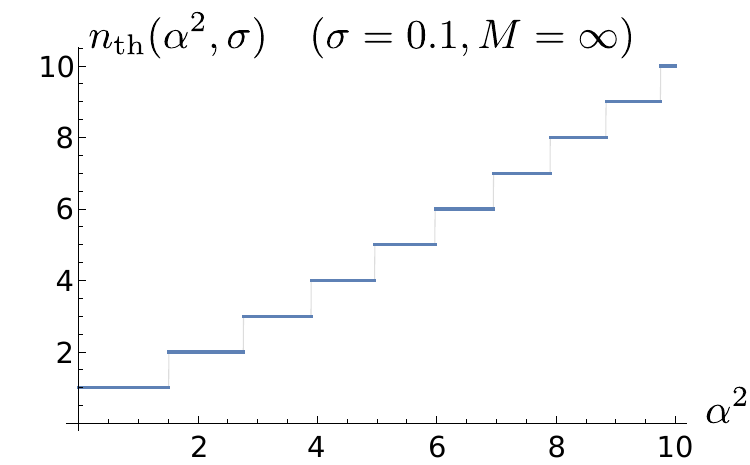} 
}
\caption{(Left) Error probability $P^{(M)}_\disp$ of the DPNR receiver as a function of the signal energy $\alpha^2$ for different photon number resolution $M$. The PNR$(1)$ case corresponds to the Kennedy receiver.  (Right) Threshold count $\nth$ as a function of the signal energy $\alpha^2$ for PNR$(\infty$) detectors. In both the pictures we fix the noise value to $\sigma=0.1$.}\label{fig:03-DPNR-vs-alpha}
\end{figure}

Plots of $P^{(M)}_\disp$ as a function of the signal energy $\alpha^2$ are reported in the left panel of Fig.~\ref{fig:03-DPNR-vs-alpha} for the realistic noise value $\sigma=0.1$ \cite{Chin2021}.
The error probability is not a monotonic function of $\alpha^2$ and, as demonstrated in Ref.~\citen{Olivares2013}, the Kennedy receiver is not near-optimum anymore in the presence of noise. 
The Kennedy is beaten by DPNR receivers with higher resolution $M$, whose corresponding error probabilities exhibit a step-like behaviour. This follows from the application of MAP criterion: as displayed in Fig.~\ref{fig:03-DPNR-vs-alpha} (right panel), for $\alpha^2\ll 1$ the threshold decision count is equal to $\nth=1$, equivalent to on-off detection, whilst for larger $\alpha^2$ it jumps to higher integer values, until reaching $\nth=M$ in the high-energy limit $\alpha^2\gg 1$. 
Accordingly, the error probability has a cusp at every change in the value of $\nth$ and, once $\nth=M$, it becomes an increasing function of the energy.
In fact, in the high-energy limit a decision error occurs only when outcome $n=M$ is retrieved from state $\rho_0$, therefore the error probability is equal to \cite{Notarnicola2023-HY, Notarnicola2023-HFF}:
\begin{align}
P^{(M)}_\disp \approx \frac{p_\sigma(M|0)}{2}= \frac12 \left[1-\sum_{j=0}^{M-1} \int_{\mathbb{R}} d\phi \, g_\sigma(\phi) \, e^{-\mu_0(\phi)}\frac{\mu_0(\phi)^j}{j!}\right] \, ,
\end{align}
being an increasing function of $\alpha^2$.
On the contrary, PNR$(\inf)$ detectors do not have a finite resolution, therefore $\nth$ can get arbitrary large values and the step-like behaviour is observed in every energy regime, as shown in Fig.~\ref{fig:03-DPNR-vs-alpha} (left panel).

\begin{figure}[tb!]
\centerline{\includegraphics[width=0.5\textwidth]{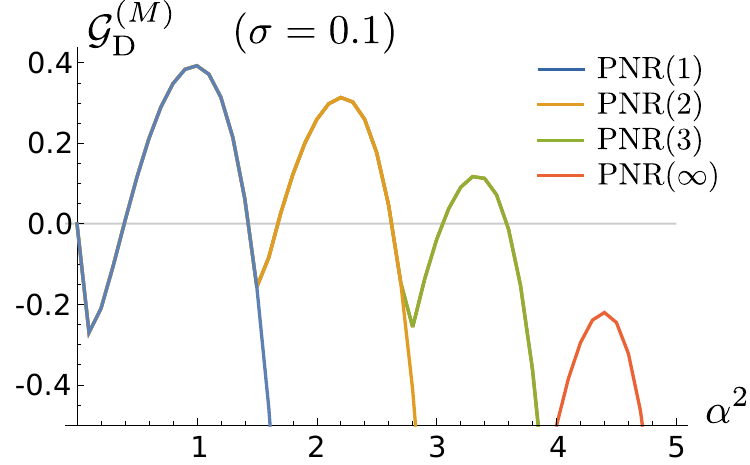}}
\caption{Gain ${\cal G}^{(M)}_\disp$ of the DPNR receiver with respect to the SQL as a function of the signal energy $\alpha^2$ for different photon number resolution $M$, when ${\cal G}^{(M)}_\disp > 0$ we beat the SQL. The PNR$(1)$ case corresponds to the Kennedy receiver. We fix the noise value to $\sigma=0.1$.}\label{fig:04-DPNR-Gain}
\end{figure}

Differently from the noiseless case, the SQL is beaten by DPNR receivers only in particular energy regimes. To highlight this, we consider the gain
\begin{align}
{\cal G}_\disp^{(M)}= 1 - \frac{P^{(M)}_\disp}{P_\SQL} \, ,
\end{align}
plotted in Fig.~\ref{fig:04-DPNR-Gain}. In turn, we have a genuine quantum advantage over the SQL when ${\cal G}_\disp^{(M)} > 0$. As expected, the gain ${\cal G}_\disp^{(M)}$ is not monotonic with $\alpha^2$, but it exhibits $M$ jumps before decreasing monotonously. The DPNR receiver outperforms the SQL in the low-energy limit and only in particular intervals of $\alpha^2$. Remarkably, for a given noise $\sigma$ we obtain the maximal region of positive gain with PNR$(M)$ detectors having sufficiently small $M$, whereas increasing the resolution further is not necessary to enhance the violation of the SQL.

\begin{figure}[tb!]
\centerline{
\includegraphics[width=0.45\textwidth]{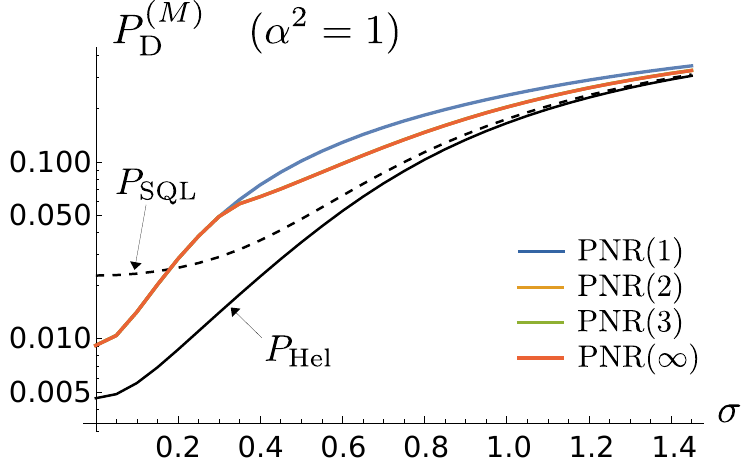} \quad
\includegraphics[width=0.45\textwidth]{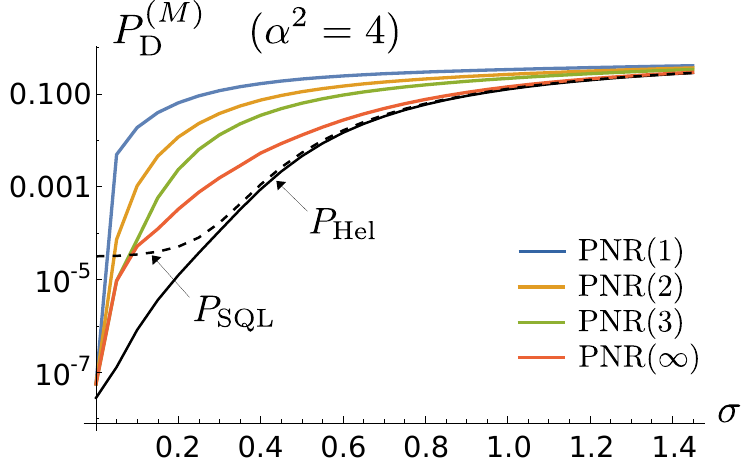} 
}
\caption{Error probability $P^{(M)}_\disp$ of the DPNR receiver as a function of the noise $\sigma$ for $\alpha^2=1$ (left) and $\alpha^2=4$ (right). For $\alpha^2=1$, the curves of PNR$(M)$ detection with $M \ge 2$ are superimposed and, thus, indistinguishable. Given the energy $\alpha^2$, the DPNR receiver outperforms the SQL in the small-noise regime, whereas for large noise the SQL becomes near-optimum.}\label{fig:05-DPNR-vs-sigma}
\end{figure}

In Fig.~\ref{fig:05-DPNR-vs-sigma} we report the error probability $P_\disp^{(M)}$ as a function of the noise $\sigma$ for low and high energy values $\alpha^2=1$ (left panel) and $\alpha^2=4$ (right panel), respectively. In both the cases DPNR receivers beat the SQL only for small noise, whilst in the large-noise limit the SQL becomes near optimum \cite{Olivares2013}. We also note that for $\alpha^2=1$ the performance of PNR$(M)$ detectors with $M\ge2$ is the same, since in this case the threshold count is equal to $\nth=2$. On the contrary, for $\alpha^2=4$ increasing the PNR resolution is beneficial to reduce the error probability.

\begin{figure}[tb!]
\centerline{\includegraphics[width=0.5\textwidth]{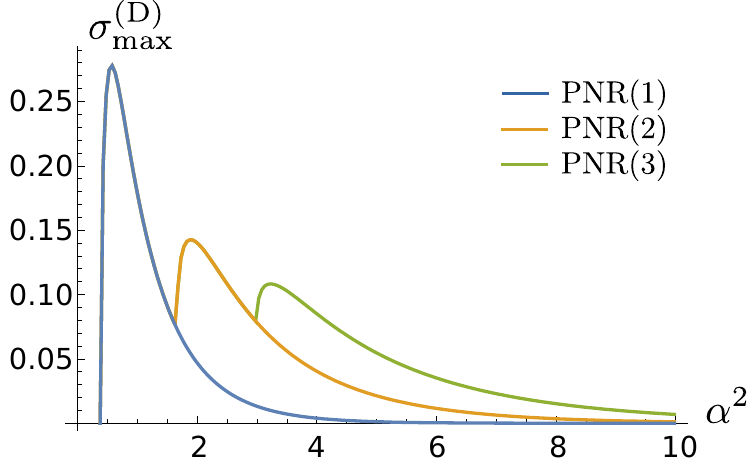}}
\caption{Maximum tolerable phase noise $\sigmamax^{(\disp)}$ as a function of the signal energy $\alpha^2$ for different photon number resolution $M$. The DPNR receiver beats the SQL in the undergraph region, namely $\sigma < \sigmamax^{(\disp)}$.}\label{fig:06-DPNR-sigmaMAX}
\end{figure}

Given the previous considerations, we introduce as a figure of merit the maximum tolerable phase noise $\sigmamax^{(\disp)}$, namely the maximum level of noise for which ${\cal G}_\disp^{(M)}\ge 0$ for a given signal energy $\alpha^2$, depicted in Fig.~\ref{fig:06-DPNR-sigmaMAX}. Thus, the DPNR receiver outperforms the SQL if $\sigma < \sigmamax^{(\disp)}$, corresponding to the undergraph region of $\sigmamax^{(\disp)}$. We have $\sigmamax^{(\disp)}=0$ for $\alpha^2<\alpha^2_\disp$, since in that regime the DPNR does not beat the SQL neither in the noiseless case; then the plot exhibits $M$ peaks and, thereafter, it decreases towards $0$.

\section{HYNORE in the presence of phase diffusion}\label{sec: HYNORE}
Now, we address the role of the HYNORE for BPSK discrimination of phase-diffused coherent states. As discussed in the former section, quantum receivers based on either quadrature measurements or displacement and photon counting show different degrees of robustness against phase noise, therefore a hybrid scheme like the HYNORE, based on the combination of both of them, provides a good candidate to better mitigate the impact of the noise.

To evaluate the performance of the HYNORE we proceed as follows. After the beam splitter with transmissivity $\tau$, the dephased signal $\rho_{k}$ is split into the separable bipartite state
\begin{align}
    \Xi_{k} = \int_\mathbb{R} d\phi\, g_\sigma(\phi)  \left|\alpha_k^{(r)} e^{-i \phi} \right\rangle \left\langle \alpha_k^{(r)} e^{-i \phi}\right| \otimes \left|\alpha_k^{(t)} e^{-i \phi} \right\rangle \left\langle \alpha_k^{(t)} e^{-i \phi} \right| \, ,
\end{align}
with $\alpha_k^{(r)}$ and $\alpha_k^{(t)}$ introduced in Eq.~(\ref{eq:alphaBS}).
Then, we perform HL detection on the first branch obtaining outcome $\Delta$ and displace the conditional state on the second branch accordingly, obtaining the (not normalized) state $\OUTPUT_k(\Delta)= \Tr_r[U \, \Xi_{k} \, U^{\dag}]$, where the reflected beam has been traced out, and
$$U=\mathbb{P}_\Delta\otimes \big\{\Theta(\Delta) D(\sqrt{\tau}\alpha)+[1-\Theta(\Delta)] D(-\sqrt{\tau}\alpha)\big\},$$
$\mathbb{P}_\Delta$ being the projection operator over the eigenspace associated with the outcome $\Delta$ and $\Theta(\Delta)$ is the Heaviside Theta function, returning $1$ for $\Delta\ge 0$ and $0$ elsewhere.
In turn, we have:
\begin{align}
\OUTPUT_k(\Delta) =& \int_\mathbb{R}  d\phi\, g_\sigma(\phi) \, {\cal S}(\Delta|\alpha_k^{(r)} e^{-i \phi}) \nonumber \\
& \hspace{1.cm} \times \Bigg\{ \Theta(\Delta) \left|\sqrt{\mu_k(\tau\alpha^2,\phi)} \, e^{-i\phi/2}\right\rangle \left\langle \sqrt{\mu_k(\tau\alpha^2,\phi)} \, e^{-i\phi/2} \right| \nonumber \\
& \hspace{1.cm} + [1-\Theta(\Delta)] \left|\sqrt{\mu_{k \oplus 1}(\tau\alpha^2,\phi)} \, e^{-i\phi/2}\right\rangle \left\langle \sqrt{\mu_{k \oplus 1}(\tau\alpha^2,\phi)} \, e^{-i\phi/2} \right|\Bigg\} \, ,
\end{align}
${\cal S}(\Delta|\alpha_k^{(r)} e^{-i \phi})$ being the HL probability of Eq.~(\ref{eq:HL}) and ``$\oplus$'' denoting the mod~$2$ sum.
Finally, we implement PNR($M$) detection on states $\OUTPUT_k(\Delta)$. The resulting joint probability of outcomes $-M\le \Delta\le M$ and $n=0,\ldots,M$ reads:
\begin{align}
p_\sigma(\Delta,n|k)&= \int_\mathbb{R}  d\phi\, g_\sigma(\phi) \, {\cal S}(\Delta|\alpha_k^{(r)} e^{-i \phi}) \nonumber \\
& \hspace{1.0cm} \times \bigg\{ \Theta(\Delta) q_n\left(\mu_k(\tau\alpha^2,\phi)\right) + [1-\Theta(\Delta)] q_n\left(\mu_{k \oplus 1}(\tau\alpha^2,\phi) \right) \bigg\} \, .
\end{align}

We perform discrimination according to the MAP criterion, i.e. by considering the threshold count depicted in Fig.~\ref{fig:03-DPNR-vs-alpha} (right panel), with the remark that the energy value to be considered is now the transmitted fraction $\tau \alpha^2$, namely $\nth=\nth(\tau\alpha^2,\sigma)$. Accordingly, the decision rule is modified as in Table \ref{Tab:02-PhN}.

\begin{table}[ph]
\tbl{Decision strategy for the HYNORE in the presence of phase diffusion.\label{Tab:02-PhN}}
{\begin{tabular}{@{}ccc@{}} \toprule
outcomes & & decision \\  \colrule
    $\Delta \ge 0$ $\quad n<\nth$\, & & ``0" \\ 
    $\Delta < 0$ $\quad n\ge\nth$ \, & & ``0" \\ 
    $\Delta < 0$ $\quad n<\nth$ \, & & ``1" \\ 
    $\Delta \ge 0$ $\quad n\ge\nth$ \, & & ``1" \\ \botrule
\end{tabular}}
\end{table}

The error probability is obtained as
\begin{align}\label{eq: HY-PhN}
P^{(M)}_\HY = \min_{\tau,z} P^{(M)}_\HY(\tau,z) \, ,
\end{align}
where
\begin{align}
    P^{(M)}_\HY(\tau,z) &= \frac12 \big[p_\sigma(\Delta < 0, n<\nth | 0) + p_\sigma(\Delta \ge 0, n\ge \nth| 0) \nonumber \\
&\hspace{1.cm} +p_\sigma(\Delta < 0, n\ge\nth| 1) + p_\sigma(\Delta \ge 0, n<\nth | 1) \big] \nonumber \\
    &= \frac12 \int_\mathbb{R} d\phi g_\sigma(\phi) \ \Biggl\{ \sum_{n=0}^{\nth-1} q_n\left(\mu_1(\tau \alpha^2,\phi) \right) \nonumber \\
&\hspace{1.cm} \times \Biggl[  \sum_{\Delta=-M}^{-1} \mathcal{S}(\Delta|\alpha_0^{(r)} e^{-i \phi}) + \sum_{\Delta=0}^{M} \mathcal{S}(\Delta|\alpha_1^{(r)} e^{-i \phi}) \Biggr]\nonumber \\
    & \hspace{1.5cm}  + \sum_{n=\nth}^{M} q_n\left(\mu_0(\tau \alpha^2,\phi) \right) \nonumber \\
&\hspace{2.cm} \times \Biggl[  \sum_{\Delta=-M}^{-1} \mathcal{S}(\Delta|\alpha_1^{(r)} e^{-i \phi}) + \sum_{\Delta=0}^{M} \mathcal{S}(\Delta|\alpha_0^{(r)} e^{-i \phi}) \Biggr] \Biggr\} \ .
\end{align}

\begin{figure}[tb!]
\centerline{
\includegraphics[width=0.5\textwidth]{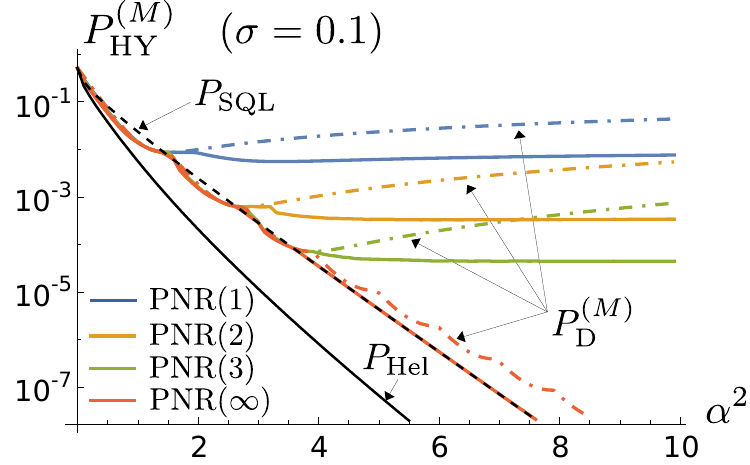}
\includegraphics[width=0.5\textwidth]{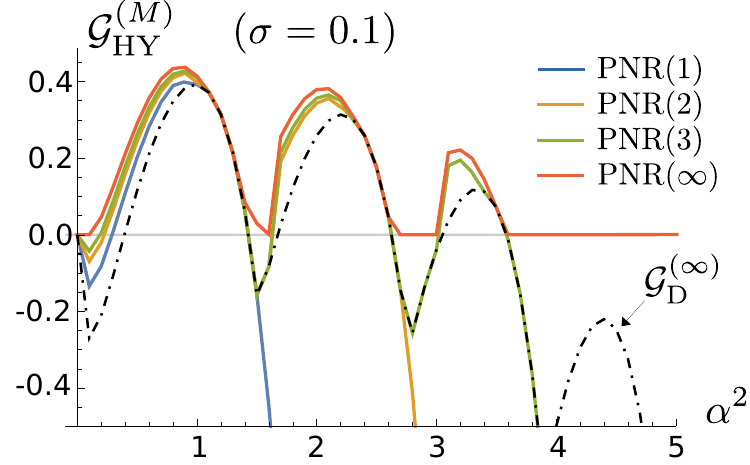} 
}
\caption{(Left) Error probability $P^{(M)}_\HY$ of the HYNORE as a function of the signal energy $\alpha^2$ for different photon number resolution $M$. The dot-dashed lines are the error probabilities of the DPNRM receiver. (Right) Gain ${\cal G}^{(M)}_\HY$ of the HYNORE with respect to the SQL as a function of the signal energy $\alpha^2$. In both the pictures we fix the noise value to $\sigma=0.1$.}\label{fig:07-HYNORE-vs-alpha}
\end{figure}

Plots of $P^{(M)}_\HY$ are reported in Fig.~\ref{fig:07-HYNORE-vs-alpha} (left panel). Like the DPNR receiver, the error probability $P^{(M)}_\HY$ exhibits a step-like behaviour induced by the change in the threshold $\nth$. The HYNORE outperforms the DPNR, $P^{(M)}_\HY \le P^{(M)}_\disp$, especially in the high-energy limit $\alpha^2\gg 1$, where the error probability is reduced of a factor $\approx 5,15,20$ for $M=1,2,3$, respectively, showing higher robustness in mitigating the phase noise.
Moreover, we observe a quantum advantage also in the low-energy regime, as emerges by computing the gain
\begin{align}
{\cal G}_\HY^{(M)}= 1 - \frac{P^{(M)}_\HY}{P_\SQL} \, ,
\end{align}
depicted in the right panel of Fig.~\ref{fig:07-HYNORE-vs-alpha}. We have ${\cal G}_\HY^{(M)} \ge {\cal G}_\disp^{(M)}$ and, differently from the DPNR case, improving the resolution $M$ makes the gain increase, since the HL scheme performs better and better, coming closer to the homodyne limit. As one may expect, the best performance is obtained with PNR$(\inf)$ detectors, where the HL performs as standard homodyne detection and ${\cal G}_\HY^{(M)} \ge 0$ for all energies.

\begin{figure}[b!]
\centerline{
\includegraphics[width=0.5\textwidth]{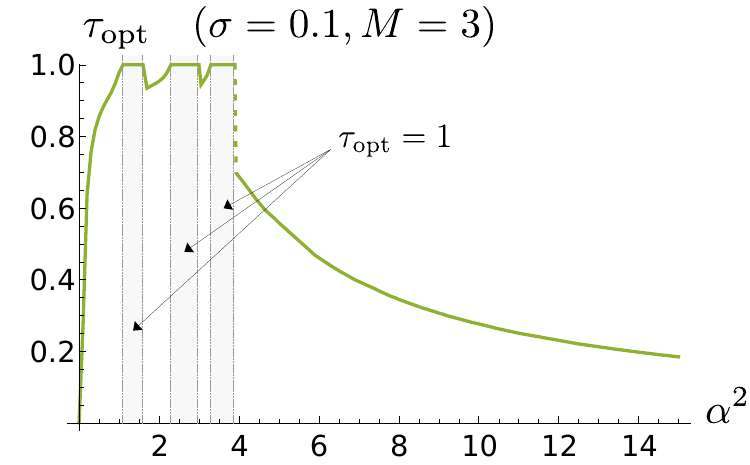}
\includegraphics[width=0.5\textwidth]{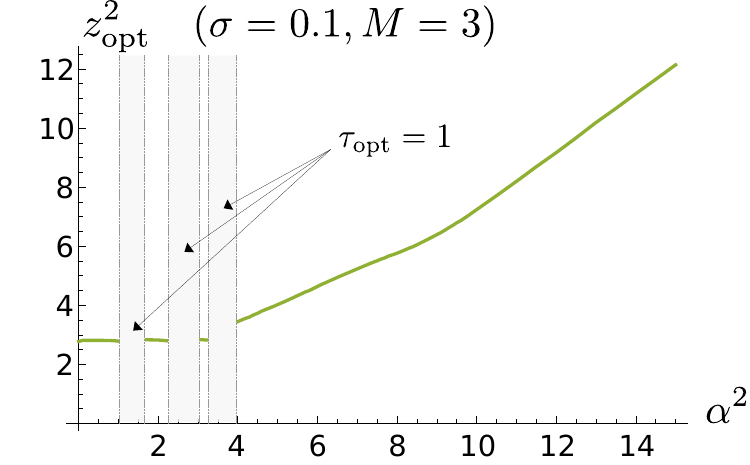} 
}
\caption{Optimized transmissivity $\tau_\opt$ (left) and LO $z^2_\opt$ (right) as a function of the signal energy $\alpha^2$ for PNR$(3)$ detectors. Both the quantities have been obtained by numerical optimization. In the shaded regions we have $\tau_\opt=1$ and the HYNORE performs as a DPNR receiver.  We fix the noise value to $\sigma=0.1$.}\label{fig:08-OptPar}
\end{figure}

The physical meaning of the present results is clearer when considering the optimized transmissivity $\tau_\opt$ and LO amplitude $z^2_\opt$ obtained after the minimization in Eq.~(\ref{eq: HY-PhN}), reported in the left and right panels of Fig.~\ref{fig:08-OptPar}, respectively, for the case of PNR$(3)$ detectors. Analogous results can be retrieved for other values of the resolution $M$.
In the low-energy limit, the transmissivity $\tau_\opt$ increases with $\alpha^2$ up to reach $1$ (corresponding to DPNR). Thereafter, we observe $M-1$ ``sawteeth", namely regions where $\tau_\opt<1$ before increasing to again reach $1$. Accordingly, when $\tau_\opt<1$ the LO is $z^2_\opt \approx M$ and the HYNORE outperforms the DPNR, whilst when $\tau_\opt=1$ all signal is sent to the transmitted DPNR setup.
On the contrary, in the high-energy limit the transmissivity jumps discontinuously and becomes a decreasing function of $\alpha^2$, saturating for $\alpha^2\gg 1$ to an asymptotic value $\tau_\inf \ne 0$. Remarkably, $\tau_\inf<1$, therefore the optimal strategy is obtained with a proper combination of both the HL and the DPNR schemes. In this regime, $z^2_\opt$ increases with $\alpha^2$, being a linear function for $\alpha^2 \gg 1$.

\begin{figure}[tb!]
\centerline{
\includegraphics[width=0.45\textwidth]{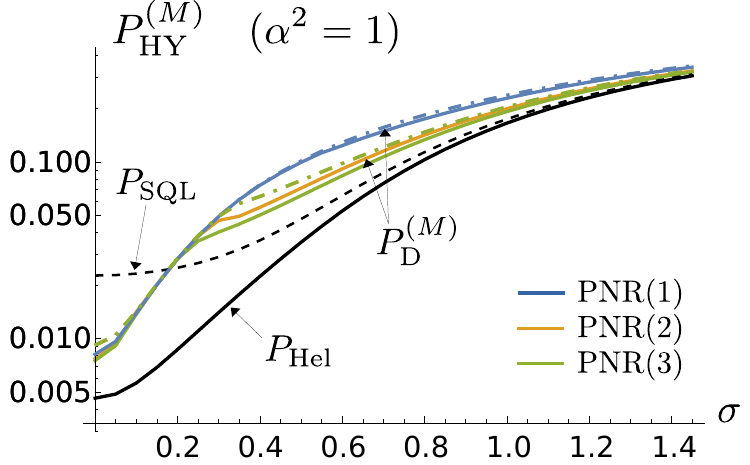} \quad
\includegraphics[width=0.45\textwidth]{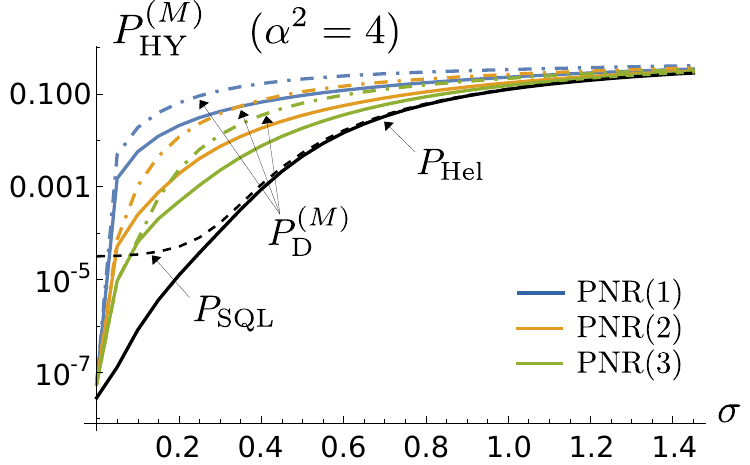} 
}
\caption{Error probability $P^{(M)}_\HY$ of the DPNR receiver as a function of the noise $\sigma$ for $\alpha^2=1$ (left) and $\alpha^2=4$ (right), compared to the DPNR receiver.}\label{fig:09-HY-vs-sigma}
\end{figure}

\begin{figure}[tb!]
\centerline{\includegraphics[width=0.5\textwidth]{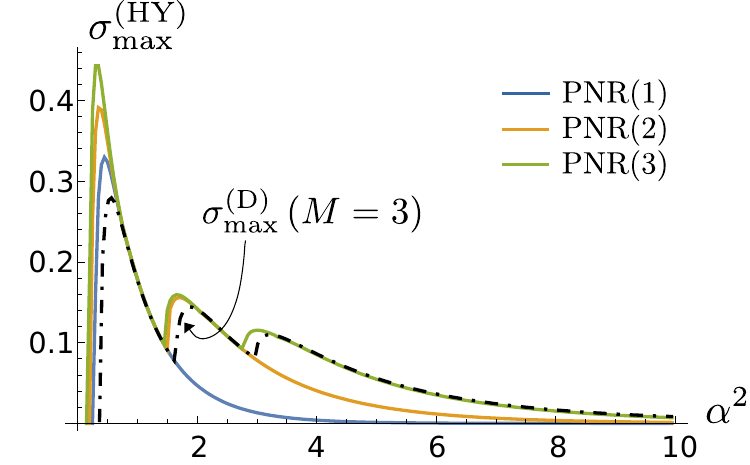}}
\caption{Maximum tolerable phase noise $\sigmamax^{(\HY)}$ as a function of the signal energy $\alpha^2$ for different photon number resolution $M$. The HYNORE receiver beats the SQL in the undergraph region, namely $\sigma < \sigmamax^{(\HY)}$. The dot-dashed line refers to $\sigmamax^{(\disp)}$ for $M=3$.}\label{fig:10-HY-sigmaMAX}
\end{figure}

Finally, in Fig.~\ref{fig:09-HY-vs-sigma} we plot $P_\HY^{(M)}$ (solid lines) as a function of the noise $\sigma$ for $\alpha^2=1$ (left panel) and $\alpha^2=4$ (right panel), respectively, comparing it to the DPNR (dot-dashed lines). We see that $P_\HY^{(M)}  \le P_\disp^{(M)}$ in both the small- and large-noise limits and the enhancement is more relevant for large $\alpha^2$, consistently with the previous analysis. As a consequence, the HYNORE increases the maximum tolerable phase noise $\sigmamax^{(\HY)}$, as depicted in Fig.~\ref{fig:10-HY-sigmaMAX}. In fact, we have $\sigmamax^{(\HY)}\ge \sigmamax^{(\disp)}$ for all energies, and $\sigmamax^{(\HY)}=0$ for $\alpha^2<\alpha^2_\HY(M)$, enlarging the region of quantum advantage with respect to the DPNR. Moreover, increasing the resolution $M$ lets the height of the peaks increase, improving further the robustness of the receiver. 

\section{Conclusions}\label{sec: Concl}

In this paper we addressed the problem of BPSK discrimination in the presence of phase diffusion noise, and investigate the performance of the HYNORE \cite{Notarnicola2023-HY, Notarnicola2023-HFF} with the intent of mitigating the noise impact. We showed that HYNORE outperforms the DPNR receiver \cite{DiMario2018, DiMario2019}, proving itself as a robust receiver to counteract phase diffusion in both the low- and high-energy limits. In particular, in the low-energy regime the HYNORE beats the SQL, therefore being a feasible solution to obtain a quantum advantage even in realistic conditions.
Moreover, we introduced the concept of maximum tolerable phase noise $\sigmamax$ to quantify the robustness of a quantum receiver, proving that $\sigmamax^{(\HY)} \ge \sigmamax^{(\disp)}$.

The results obtained in the paper provide a more complete characterization of feasible quantum receivers, identifying the limits of quantum communications in the presence of realistic non-Gaussian noise. Furthermore, they pave the way for new applications in quantum phase communications\cite{Hall91,Jarzyna2016,Adnane2019} and continuous-variable QKD\cite{Grosshans02,Grosshans07,Notarnicola2024-SEC}.



\renewcommand\bibname{References}


\begin{thebibliography}{10}

\bibitem{Agrawal2002}
G.~P. Agrawal, 
{\em Fiber-optic communications systems\/} (Wiley, New York, 2002).
%
\bibitem{Kikuchi2016}
K. Kikuchi, 
{\em J. Light. Technol.\/} {\bf 34} (2016) 157--179.
%
\bibitem{Kaushal2017}
H. Kaushal and G. Kaddoum, 
{\em IEEE Commun. Surv. Tutor.\/} {\bf 19} (2017) 57--96.
%
\bibitem{Helstrom1976}
C.~W. Helstrom,
{\em Quantum Detection and Estimation Theory\/} (Academic Press, New York, 1976).
%
\bibitem{Bergou2010}
J.~A. Bergou, 
{\em J. Mod. Opt.\/} {\bf 57} (2010) 160--180.
%
\bibitem{Cariolaro2015}
G. Cariolaro, 
{\em Quantum Communications\/} (Springer, Berlin,
  2015). 
%
\bibitem{Giovannetti2004}
V. Giovannetti {\em et al.}, 
{\em Phys. Rev. Lett.\/} {\bf 92} (2004) 027902.
%
\bibitem{Arrazola2014}
J.~M. Arrazola and N. L\"utkenhaus,
{\em Phys. Rev. A\/} {\bf 90} (2014) 042335.
%
\bibitem{Grosshans2002}
F. Grosshans and P. Grangier, 
{\em Phys. Rev. Lett.\/} {\bf 88} (2009) 057902.
%
\bibitem{Gisin2002}
N. Gisin, G. Ribordy, W. Tittel and H. Zbinden, 
{\em Rev. Mod. Phys.\/} {\bf 74} (2002) 145--195.
%
\bibitem{Leverrier2009}
A. Leverrier and P. Grangier, 
{\em Phys. Rev. Lett.\/} {\bf 102} (2009) 180504.
%
\bibitem{Notarnicola2023-Pol}
M.~N. Notarnicola, M. Jarzyna, S. Olivares and K. Banaszek, 
{\em New J. Phys.\/} {\bf 25} (2023) 103014.
%
\bibitem{Helstrom1970}
C.~W. Helstrom, J.~W.~S. Liu and J.~P. Gordon, 
{\em Proc. IEEE\/} {\bf 58} (1970) 1578--1598.
%
\bibitem{Dolinar1973}
S.~J. Dolinar, 
{\em Quart. Prog. Rep.\/} {\bf 11} (1973) 115--120.
%
\bibitem{Lau2006}
C.~W. Lau {\em et al.},
in {\em Free-Space Laser Communication Technologies XVIII\/} Vol.~6105 (SPIE, 2006),144--150.
%
\bibitem{Kennedy1973}
R.~S. Kennedy,
{\em Quart. Prog. Rep.\/} {\bf 108} (1973) 219--225 .
%
\bibitem{Takeoka2008}
M.~Takeoka and M.~Sasaki, 
{\em Phys. Rev. A\/} {\bf 78} (2008) 022320.
%
\bibitem{DiMario2018}
M.~T. DiMario and F.~E. Becerra, 
{\em Phys. Rev. Lett.\/} {\bf 121} (2018) 023603.
%
\bibitem{Notarnicola2023-HY}
M.~N. Notarnicola, M.~G.~A. Paris and S. Olivares, 
{\em J. Opt. Soc. Am. B\/} {\bf 40} (2023) 705--714.
%
\bibitem{Assalini2011}
A. Assalini, N.~D. Pozza and G. Pierobon, 
{\em Phys. Rev. A\/} {\bf 84} (2011) 022342.
%
\bibitem{Sych2016}
D. Sych and G. Leuchs, 
{\em Phys. Rev. Lett.\/} {\bf 117} (2016) 200501.
%
\bibitem{Notarnicola2023-HFF}
M.~N. Notarnicola and S. Olivares, 
{\em Phys. Rev. A\/} {\bf 108} (2023) 042619.
%
\bibitem{Banerjee2007}
S. Banerjee and R. Srikanth
{\em Phys. Rev. A\/} {\bf 76} (2007) 062109.
%
\bibitem{Ishii11}
H. Ishii, K. Kasaya and H. Oohashi, {\em NTT Technical Review} {\bf 9}(3) (2011) 1.
%
\bibitem{Amiri22}
Z. Amiri, B. A. Bash and J. N\"otzel, 
in {\em 2022 IEEE Globecom Workshops (GC Wkshps)}, Rio de Janeiro, Brazil (2022)  298--303.
%
\bibitem{Bertaina23}
G. Bertaina {\em et al.}, 
Phase Noise in Real-World Twin-Field Quantum Key Distribution,
arXiv:2310.08621 [quant-ph].
%
\bibitem{Genoni2011}
M.~G. Genoni, S. Olivares and M.~G.~A. Paris, 
{\em Phys. Rev. Lett.\/} {\bf 106} (2011) 153603.
%
\bibitem{Cialdi2020}
S. Cialdi, E. Suerra, S. Olivares, S. Capra and M.~G.~A. Paris, 
{\em Phys. Rev. Lett.\/} {\bf 124} (2020) 163601.
%
\bibitem{Altorio2015}
M. Altorio, M.~G. Genoni, M.~D. Vidrighin, F. Somma and M. Barbieri,
{\em Phys. Rev. A\/} {\bf 92} (2015) 032114.
%
\bibitem{Szczykulska2017}
M. Szczykulska, T. Baumgratz and A. Datta,
{\em Quantum Sci. Technol.\/} {\bf 2} (2017) 044004 .
%
\bibitem{Notarnicola2022}
M.~N. Notarnicola, M.~G. Genoni, S. Cialdi, M.~G.~A. Paris and S. Olivares, 
{\em J. Opt. Soc. Am. B\/} {\bf 39} (2022) 1059--1067.
%
\bibitem{Genoni2012}
M.~G. Genoni, S. Olivares, D. Brivio, S. Cialdi, D. Cipriani, A. Santamato, S. Vezzoli and M.~G.~A. Paris, 
{\em Phys. Rev. A\/} {\bf 85} (2012) 043817.
%
\bibitem{Vidrighin2014}
M. Vidrighin, G. Donati, M.~G. Genoni, X.-M. Jin, W.~S. Kolthammer, M.~S. Kim, A. Datta, M. Barbieri and I.~A. Walmsley,
{\em Nat. Commun.\/} {\bf 5} (2014) 3532.
%
\bibitem{Feng2014}
X.~M. Feng, G.~R. Jin and W. Yang,
{\em Phys. Rev. A\/} {\bf 90} (2014) 013807.
%
\bibitem{Jarzyna2014}
M. Jarzyna, K. Banaszek and R. Demkowicz-Dobrzański, 
{\em J. Phys. A\/} {\bf 47} (2014) 275302.
%
\bibitem{Trapani2015}
J. Trapani, B. Teklu, S. Olivares and M.~G.~A. Paris, 
{\em Phys. Rev. A\/} {\bf 92} (2015) 012317.
%
\bibitem{Namkung2022}
M. Namkung, J.~S. Kim,
Atomic indirect measurement and robust binary quantum communication under phase-diffusion noise
arXiv:2205.05301 [quant-ph].
%
\bibitem{Olivares2013}
S. Olivares, S. Cialdi, F. Castelli and M.~G.~A. Paris, 
{\em Phys. Rev. A\/} {\bf 87} (2013) 050303(R).
%
%
\bibitem{Chesi2018}
G. Chesi, S. Olivares and M.~G.~A. Paris, 
{\em Phys. Rev. A\/} {\bf 97} (2018) 032315.
%
\bibitem{DiMario2019}
M.~T. DiMario, L. Kunz, K. Banaszek and F.~E. Becerra, 
{\em npj Quantum Inf.\/} {\bf 5} (2019) 65.
%
\bibitem{Donati2014}
G. Donati, T.~J. Bartley, X.-M. Jin, M.-D. Vidrighin, A. Datta, M. Barbieri and I.~A. Walmsley,
{\em Nat. Commun.\/} {\bf 5} (2014) 5584.
%
\bibitem{Thekkadath2020}
G.~S. Thekkadath {\em et al.}, 
{\em Phys. Rev. A\/} {\bf 101} (2020) 031801(R).
%
%
\bibitem{Olivares2021}
S. Olivares,
{\em Phys. Lett. A\/} {\bf 418} (2021) 127720.
%
\bibitem{Paris1996}
M.~G.~A.~Paris,
{\em Phys. Lett. A\/} {\bf 217} (1996) 78--80.
%
\bibitem{Ip2008}
E. Ip, A.~P.~T. Lau, D.~J.~F. Barros and J.~M. Kahn,
{\em Opt. Express\/} {\bf 16} (2008) 753--791.
%
\bibitem{Marie2017}
A. Marie and R. Alléaume,
{\em Phys. Rev. A\/} {\bf 95} (2017) 012316.
%
\bibitem{Chin2021}
H.-M. Chin, N. Jain, D. Zibar, U.~L. Andersen and T. Gehring,
{\em npj Quantum Inf.\/} {\bf 7} (2021) 20.
%
\bibitem{Shao2021}
Y. Shao {\em et al.},
{\em Phys. Rev. A\/} {\bf 104} (2021) 032608.
%
\bibitem{Bina2016}
M. Bina, A. Allevi, M. Bondani and S. Olivares,
{\em Sci. Rep.\/} {\bf 6} (2016) 26025.
%
%
\bibitem{Hall91}
M.~J.~W. Hall and I.~G. Fuss,
{\em Quantum Opt.\/} {\bf 3} (1991) 147--167.
%
\bibitem{Jarzyna2016}
M. Jarzyna, V. Lipi\'nska, A. Klimek, K. Banaszek and M. G. A. Paris, 
{\em Opt. Express\/} {\bf 24} (2016) 1693--1698. 
%
\bibitem{Adnane2019}
H. Adnane, B. Teklu and M.~G.~A. Paris, 
{\em J. Opt. Soc. Am. B\/} {\bf 36} (2019) 2938--2945. 
%
\bibitem{Grosshans02}
F. Grosshans and P. Grangier, 
{\em Phys. Rev. Lett.\/} {\bf 88} (2002) 057902.
%
\bibitem{Grosshans07}
F. Grosshans, A. Ac\'in and N. Cerf,
in {\em Quantum Information With Continuous Variables of
Atoms and Light} (World Scientific, Singapore, 2007) 63--83.
%
\bibitem{Notarnicola2024-SEC}
M.~N. Notarnicola, S. Olivares, E. Forestieri, E. Parente, L. Potì and M. Secondini, 
{\em IEEE Trans. Commun.\/} {\bf 72} (2024) 375--386.
%



\end{thebibliography}
\end{document}